\documentclass[preprint]{vgtc}               %

\graphicspath{{figures/}{pictures/}{images/}{./}} %

\usepackage{times}                     %

\usepackage{tabu}                      %
\usepackage{booktabs}                  %
\usepackage{lipsum}                    %
\usepackage{mwe}                       %

\usepackage{mathptmx}                  %
\usepackage{booktabs}
\usepackage{enumitem} %

\onlineid{1205}

\vgtccategory{Empirical Study}
\vgtcinsertpkg

\title{Towards Effective Generation of Interactive Visualizations with Vibe Coding: An Empirical Study}

\author{Yanshan Zeng\thanks{e-mail: zengyanshan8231@gmail.com} %
\and Ruixuan Tu\thanks{e-mail: runxuantu@bit.edu.cn} 
\and Zuo Xiang\thanks{e-mail: zuoxiang@bit.edu.cn} 
\and Lijia Feng\thanks{e-mail: lijiafeng@bit.edu.cn} 
\and Guozheng Li\thanks{e-mail: guozheng.li@bit.edu.cn}
\and Chi Harold Liu\thanks{e-mail: chiliu@bit.edu.cn}}
\affiliation{\scriptsize Beijing Institute of Technology}

\abstract{
   Constructing interactive visualizations has traditionally required substantial human effort, involving both technical implementation and design decision-making. 
   Recently, vibe coding, a programming paradigm leveraging Large Language Models to generate, interpret, and refactor code from natural language specifications, has emerged as a promising approach to reduce the burden. 
   However, the capabilities and limitations of vibe coding in building interactive visualizations remain unexplored. 
   To address this gap, we conducted a user study with 78 participants that were tasked with constructing interactive visualizations using vibe coding. 
   We further collected users feedback through questionnaires, interviews, and case analyses. 
   Based on this study, we examine (1) the capabilities and (2) user experience of vibe coding in generating interactive visualizations, and (3) the practical human-agent collaboration strategies adopted. 
   Our findings provide the first systematic assessment of vibe coding for interactive visualization construction, revealing both its strengths and limitations, explaining the shift in developer labor and identifying the hybrid collaboration strategies participants adopted. 
   Furthermore, our study offers insights for more intuitive and robust vibe coding practices.
} %

\keywords{Visualization construction, vibe coding.}

\begin{document}
\maketitle

\section{Introduction}
The implementation of interactive visualization traditionally requires substantial programming effort. 
Early approaches relied on writing C++ code, and later, with the emergence of web-based frameworks like D3~\cite{D3tvcg2011} and Vega-Lite~\cite{satyanarayan2016vega}, developers shifted to web coding to implement interactive visualizations.
Despite these advances, building interactive visualizations remains time-consuming and technically demanding.
Recently, the rise of Large Language Models (LLMs) has enabled a new programming paradigm known as vibe coding, coined by Andrej Karpathy in February 2025~\cite{propose-vibe-coding:2025:x}. 
Vibe coding provides a fluid and natural interface to a high-cost environment. More specifically, developers specify goals, constraints, and modifications in natural language, while AI agents generate, interpret, and refactor the code on their behalf~\cite{vibe-coding-survey:2025:arxiv}. 
This approach promises to eliminate much of the programming burden and accelerate the creation of interactive visualizations~\cite{deepvis:2026:tvcg, smartmlvs:2025:tvcg, chartgpt:2025:tvcg, dynavis:2024:chi, directgpt:2024:chi}.

However, interactive visualization construction poses unique challenges for vibe coding. Unlike general programming tasks, interactive visualizations require the implementation of multiple interdependent components, such as data processing, visual mapping, interaction logic, and layout organization~\cite{nested-model:2009:tvcg, vega-lite:2017:tvcg, declarative:2014:uist}, and the design decisions about how data are transformed, presented, and interacted with. 
This combination of programming complexity and design reasoning makes interactive visualization construction a particularly challenging scenario for vibe coding.

To understand how far we are from realizing end-to-end interactive visualization generation, we conducted a user study in university with 78 students tasked with constructing interactive visualizations while predominantly using vibe coding. 
Through questionnaires, interviews, and case analyzes, we examined the current capabilities, limitations, and strategies associated with vibe coding in interactive visualization construction.
Based on this study, our goal was to answer three research questions (RQs).
\begin{itemize}[leftmargin=*]
\item RQ1 (Capability): How effective is vibe coding for interactive visualization construction and how does performance vary across task complexity?
\item RQ2 (User Experience): What is the user experience when using vibe coding for interactive visualization construction?
\item RQ3 (Strategy): What strategies do developers adopt in vibe coding for interactive visualization construction?
\end{itemize}

Following the task, participants completed questionnaires covering delivery efficiency, code reliability, and cognitive load, and a subset participated in semi-structured interviews to investigate the underlying causes of friction and identify best practices. Our findings indicate that while vibe coding accelerates prototyping, it struggles with fine-grained interaction and visualization designs as well as system-level consistency. Participants experience a shift in labor from coding to semantic alignment, often resulting in a description-verification bottleneck. To overcome these limitations, they often employed hybrid collaboration strategies that combine high-level prompting with selective manual code intervention.

In conclusion, our contributions are as follows.
\begin{itemize}[leftmargin=*]
\item The first systematic assessment of vibe coding in interactive visualization construction.
\item Characterization of the core advantages, pain points, and interaction patterns of vibe coding across functional capabilities, user experience, and collaboration strategies.
\item Key design implications and support requirements for future vibe coding techniques in interactive visualization.
\end{itemize}

\section{Related Work}

Previous studies have examined vibe coding in general-purpose programming, focusing on efficiency, validation costs, and interaction methods~\cite{validate-ai-code:2024:chi, copilot-user-behavior:2024:chi}. In visualization, LLMs have been applied to tasks such as chart generation, multi-view construction, and visualization code synthesis~\cite{lightva:2025:tvcg, data-formulator:2024:tvcg, dynavis:2024:chi, celestial:2026:chi, llm-vis-item:2025:tvcg}.
However, most studies focus on isolated tasks or the perspective of end users, such as data analysts, rather than on developers constructing interactive visualization systems.
Human-AI collaboration is central to LLM-assisted programming, which typically involves prompting, reviewing, and verifying~\cite{al-khalifa-2026-code-centric}. Existing work has mainly addressed general programming challenges~\cite{sarkar2025vibe}. In particular, the practical capabilities and user experiences of developers using vibe coding remain insufficiently examined. Furthermore, prior work has yet to systematically investigate collaboration strategies for design decision-making and the collaborative processes involved in building complete interactive visualization systems.

Our study tries to address this gap through a user-centered evaluation of vibe coding in interactive visualization construction, providing insight into its capabilities, limitations, user experiences, and collaboration strategies.

\section{Method}

\begin{figure*}[t]
  \centering
  \includegraphics[width=\textwidth]{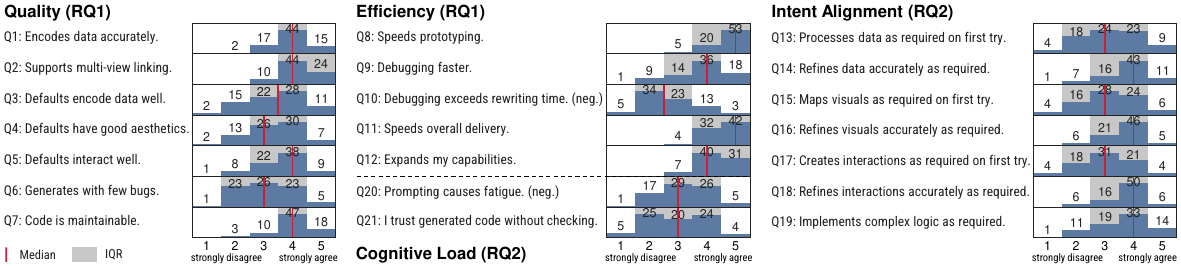}
  \caption{Results for 5-point Likert scale items in the post-task questionnaire, which are divided into four categories, quality and efficiency are for RQ1, cognitive load and intent alignment are for RQ2.}
  \label{fig:likert_item_distribution}
\end{figure*}

\subsection{Participants}

We recruited 78 participants (P1--P78) with computer science backgrounds from universities (53 males and 25 females), all of whom possessed knowledge of interactive visualizations and experience with vibe coding. Their LLM usage patterns varied: 21 relied almost entirely on LLMs, 54 used LLMs to assist their own frameworks and logic, and 3 used LLMs only for troubleshooting. The LLMs and related tools they commonly used included Codex, Claude Code, Cursor, and GitHub Copilot.

\subsection{Procedure}

The study was conducted in three stages: an interactive visualization construction task, a post-task questionnaire, and a semi-structured interview.

(1) Interactive visualization construction task (1 week). Participants were instructed to develop an interactive visualization with vibe coding, including appropriate data transformation, visual mapping, and interactions that support data exploration while ensuring a clear, user-friendly, and responsive interface. 
The participants were given the autonomy to select their own topics and datasets.

(2) Post-task questionnaire (20 minutes). After completing the development of interactive visualizations, participants completed a questionnaire to provide a retrospective self-report on the capabilities and user experience of vibe coding, as well as the human-agent collaboration strategies employed during their development.

(3) Semi-structured interview (25 minutes). To further investigate issues and best practices beyond the questionnaire, we invited participants to conduct individual semi-structured interviews.

\subsection{Measurement and Analysis}

The data collected are analyzed using a mixed-method approach.

\textbf{Quantitative analysis:} For the five-point Likert-scale questions in the questionnaire, we calculated descriptive statistics, including mean, standard deviation, median, and interquartile range (IQR), to assess participants' perceptions of both performance metrics (e.g., delivery efficiency and code reliability) and experiential factors (e.g., intent alignment and cognitive load).

\textbf{Qualitative analysis:} For open-ended questions and interview records, we conducted a thematic coding analysis. Google's Gemini~\cite{gemini:2026:google} was initially used to code the data, followed by manual review to identify recurring themes related to the advantages and limitations of vibe coding, as well as collaborative strategies.

\section{Result}

\subsection{Capabilities of Vibe Coding (RQ1)}

The results of our experiment demonstrate that vibe coding accelerates the delivery of interactive visualization systems but exhibits significant performance inconsistencies. 
Although it offers distinct advantages in prototyping and framework generation, it still requires ongoing developer intervention and refinement in terms of the precision of complex interactions, the quality of detailed visualization designs, and code stability.

Vibe coding received high rates for its overall efficiency in delivering interactive visualizations, particularly in terms of prototyping efficiency. The results indicate that the participants generally believed that vibe coding significantly reduced the time required to build initial prototypes of visualization systems (\cref{fig:likert_item_distribution}-Q8); although the systems built with vibe coding still contained a certain number of bugs (\cref{fig:likert_item_distribution}-Q6), fixing bugs in the code using vibe coding was more advantageous than manual debugging (\cref{fig:likert_item_distribution}-Q9); taking into account the time spent coding, debugging, and refactoring, vibe coding indeed accelerated the final delivery of the entire interactive visualization (\cref{fig:likert_item_distribution}-Q11). In open questions, some participants (35.9\%) mentioned that vibe coding enables the rapid setup of basic frameworks and charts, further illustrating its strong support during the ``from scratch'' development phase.

\begin{figure}[tb]
 \centering %
 \includegraphics[width=\columnwidth]{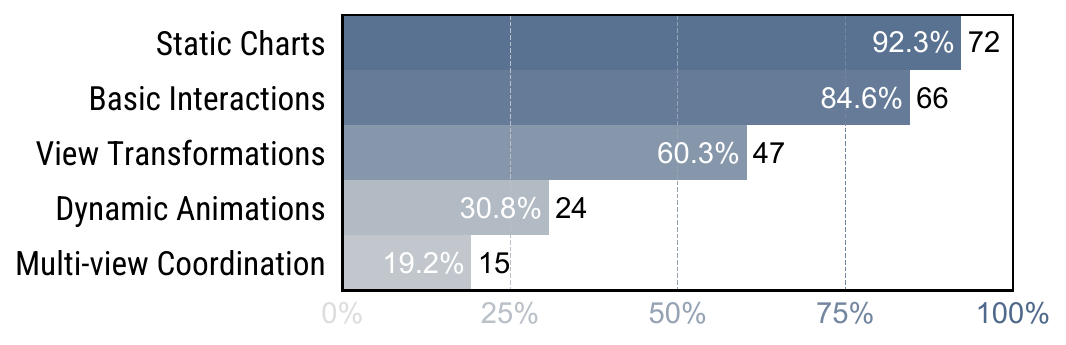}
 \caption{Respondent counts for successful one-shot generation of vibe coding across different visualization task categories (N=78). Specifically, ``Basic Interactions'' refer to simple user interface triggers such as tooltips and legend toggling; ``View Transformation'' encompasses navigation-based actions such as zooming and panning.}
 \label{fig:one_shot_success_rate}
\end{figure}

The capabilities of vibe coding are not evenly distributed across different types of interactive visualizations. 
As shown in \cref{fig:one_shot_success_rate}, participants noted that vibe coding demonstrates a high success rate on a single attempt primarily for generating static charts and basic interactions, but performs poorly when it comes to generating complex interactions and multi-view coordination. 
In particular, LLMs often forget the ``shared state'' (like a global variable for brushing/linking) when generating code for individual views in isolation.
As tasks progress from simple to complex, from generic to personalized, and from individual components to system-wide coordination, the quality and performance of vibe coding declines.

Vibe coding has limitations in terms of code reliability and generation accuracy. Although participants acknowledged that vibe coding generates code quickly and produces a large volume of output, they did not believe the code to be stable enough to be delivered directly without manual review (\cref{fig:likert_item_distribution}-Q6 and Q21). 
In open questions, participants frequently mentioned issues such as confusing interaction logic (25.6\%), mismatched state synchronization between views (12.8\%) and improper visual mapping (12.8\%). These indicate that while the codes generated by vibe coding are generally functional, they lack sufficient quality in the details.

\subsection{User Experience in Vibe Coding (RQ2)}

The user experience in vibe coding is characterized by a complex trade-off: while it lowers the barrier to entry, it introduces significant cognitive friction and mental fatigue during refinement. The imprecision of vibe coding in faithfully translating linguistic intent into visual and interactive details, coupled with the lack of predictable feedback, shifts the mental effort away from manual coding to an exhausting loop of re-prompting and validating.

The most prominent usability issue stems from the high cost of intent expression and semantic alignment, which creates a cognitive burden. The participants generally felt that the initial results generated by the vibe coding often did not match their intentions fully (\cref{fig:likert_item_distribution} -Q13, Q15, and Q17), requiring repeated communication with the agent, leading to fatigue (\cref{fig:likert_item_distribution}-Q20). When describing the difficulties they encountered, ``semantic understanding discrepancies and high barriers to prompt construction'' were cited most frequently (43.6\%). 
A notable proportion (37.2\%) of the participants complained that while modifying the prompt might improve the results, it might also introduce new biases, leaving them feeling they lack control over development. 
For interactive visualization tasks that involve constraints on layout, animation, events, and visual style, this communication cost is particularly pronounced.

The primary challenge in intent expression and semantic alignment lies in the difficulty of accurately conveying visual encoding details and complex interactive nuances through natural language.
Participants noted that their primary frustration stemmed not from the agent's inability to generate functional code but from the high interaction cost required for fine-grained refinement. 
Even when initial outputs were broadly aligned with user requirements, achieving precise visual and interactive specifications, such as layout margins, element positioning, and complex interaction sequences—necessitated exhaustive iterative prompting.

In addition to struggling to achieve the desired results, developers also find it difficult to grasp how the interface will appear and how interactions will occur after modifications. 
25.6\% of the participants mentioned that they had to constantly switch between the agent dialog and the window that displays the visual interface to observe the effects of the changes. 
This back-and-forth between natural language and visual modalities increases the cognitive load involved in interactive visual development. 
Current vibe coding tools still lack sufficiently intuitive intermediary mechanisms for linking language commands, interface previews, and interactive feedback.

\subsection{Collaboration Strategies in Vibe Coding (RQ3)}

The proficiency in vibe coding is predicated on the developer’s ability to orchestrate collaboration between multiple abstraction levels. 
This requires a strategic determination of when to impose structural constraints, using sample datasets, reference images, and incremental prompting, versus when to grant agent generative autonomy to expand system capabilities.
Ultimately, the process necessitates a persistent ``human-in-the-loop'' approach to critically evaluate outputs and ensure design alignment."

When interacting with the agent, developers do not passively wait for the agent to understand their requirements; instead, they actively simplify the task and constrain the generated results. Most participants chose to break down the process into steps and incrementally add features (61.5\%) or provide sample data to help the agent understand the structure of the data (59.0\%). Some participants opted to provide reference diagrams (32.1\%) or issue a single and lengthy instruction that contained all task details at once (28.2\%). To achieve better results with vibe coding, developers often proactively organize, decompose, and constrain tasks.

During interactive visualization development, vibe coding is not used for fully automated generation but rather as a hybrid collaborative process involving both agent-assisted and manual intervention. 
When participants used natural language to instruct the agent to modify the existing visualization code, more than half (59.0\%) still needed to manually modify 10\%–30\% of the code; only 28.2\% of participants believed that the system could work with minor adjustments. 
This indicates that while vibe coding significantly lowers the development barrier and handles some code generation and modification tasks, developers still need to take an active role in checking and correcting the details.

In practice, participants developed different hybrid intervention strategies to determine when to rely on the agent and when to use manual methods. The approaches of the participants to the modification of the code followed a pragmatic cost-benefit logic, characterized by the following top three strategies:

(1) Considering the marginal cost of efficiency (42.3\%): if a manual modification could be completed quickly, they chose to do it manually; otherwise, they chose to guide the agent to make modifications via natural language instructions;

(2) Considering the granularity of the modification (38.5\%): if only a few parameters of a single component need to be changed, a manual modification is chosen; otherwise, the agent is guided to make the modification using natural language instructions;

(3) Considering the cost of understanding and cognitive load (26.9\%): the priority is given to having the agent, which has a better grasp of the general structure and details of the code of the system, perform the modification. 

This indicates that developers can dynamically change collaboration approaches: when tasks involve extensive changes, structural adjustments, or a technology stack with which they are unfamiliar, developers are more inclined to rely on the agent; whereas when issues involve only local parameters, layout details, or simple code snippets, and the cost of continuing the dialog exceeds that of direct modification, developers opt for manual intervention.

\subsection{Use Cases}

The following cases explore the issue of homogenization in vibe coding for visualization design: the first highlights the agent's deficiency in visualization domain knowledge, while the second delineates a successful human-agent collaboration practice.

\begin{figure}[tb]
 \centering %
 \includegraphics[width=\columnwidth]{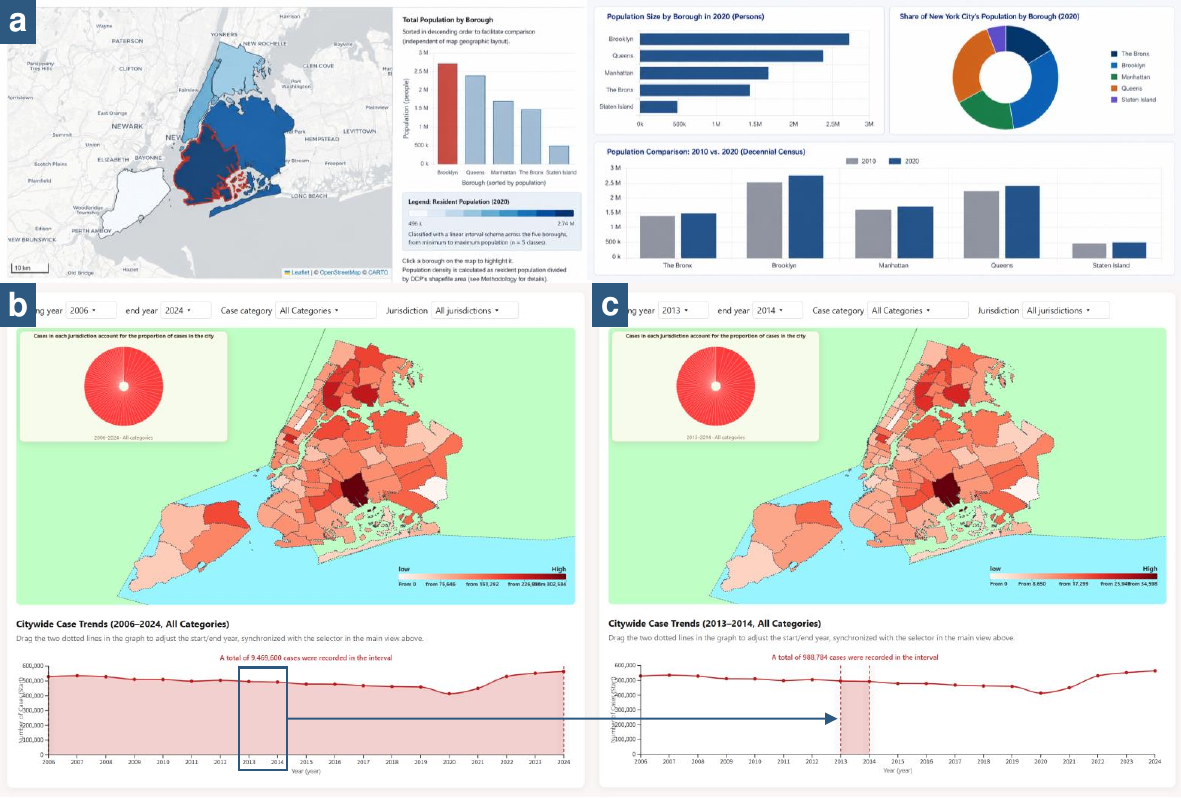}
 \caption{Case of the \textit{New York City Public Safety Landscape Dashboard} by P41. (a) Initial agent-generated dashboard under ambiguous prompts, showing generic visualizations; (b) Improved dashboard achieved through explicit design guidance and templates; (c) Misleading color encoding executed by the agent: when filtered from a broad range (2006--2024, max $\approx$ 302k) to a narrow range (2013--2014, max $\approx$ 34k), the view seems unchanged.}
 \label{fig:case_study_P41}
\end{figure}

\textbf{Agent's deficiency in visualization domain knowledge.} P41's task involved the multi-dimensional analysis of population density and crime rates in New York City. 
Initially, the agent exhibited limited design expressiveness, defaulting to basic choropleth maps and bar charts with rudimentary interactions such as zooming and tooltips (\cref{fig:case_study_P41}(a)). 
High-fidelity, information-dense visualizations were only realized through proactive human intervention, where the participant provided specific templates to bypass the agent's limited expressiveness (\cref{fig:case_study_P41}(b)).
A critical failure in the agent’s analytical reasoning emerged during the implementation of temporal filtering. 
The agent blindly executed a recalibration of the color-scale mapping based on the local data range of the filtered subset (2013–2014) rather than maintaining a global normalization context. 
This resulted (\cref{fig:case_study_P41}(c)) in a statistically misleading visualization where low-frequency data appeared as visually saturated as high-frequency global data, violating the principle of expressiveness.
This indicates that current vibe coding agents remain analytically blind to visualization principles, lacking the analytical sensitivity and design-intent reasoning required to proactively identify irrational mappings or confirm complex design.

\begin{figure}[tb]
 \centering %
 \includegraphics[width=\columnwidth]{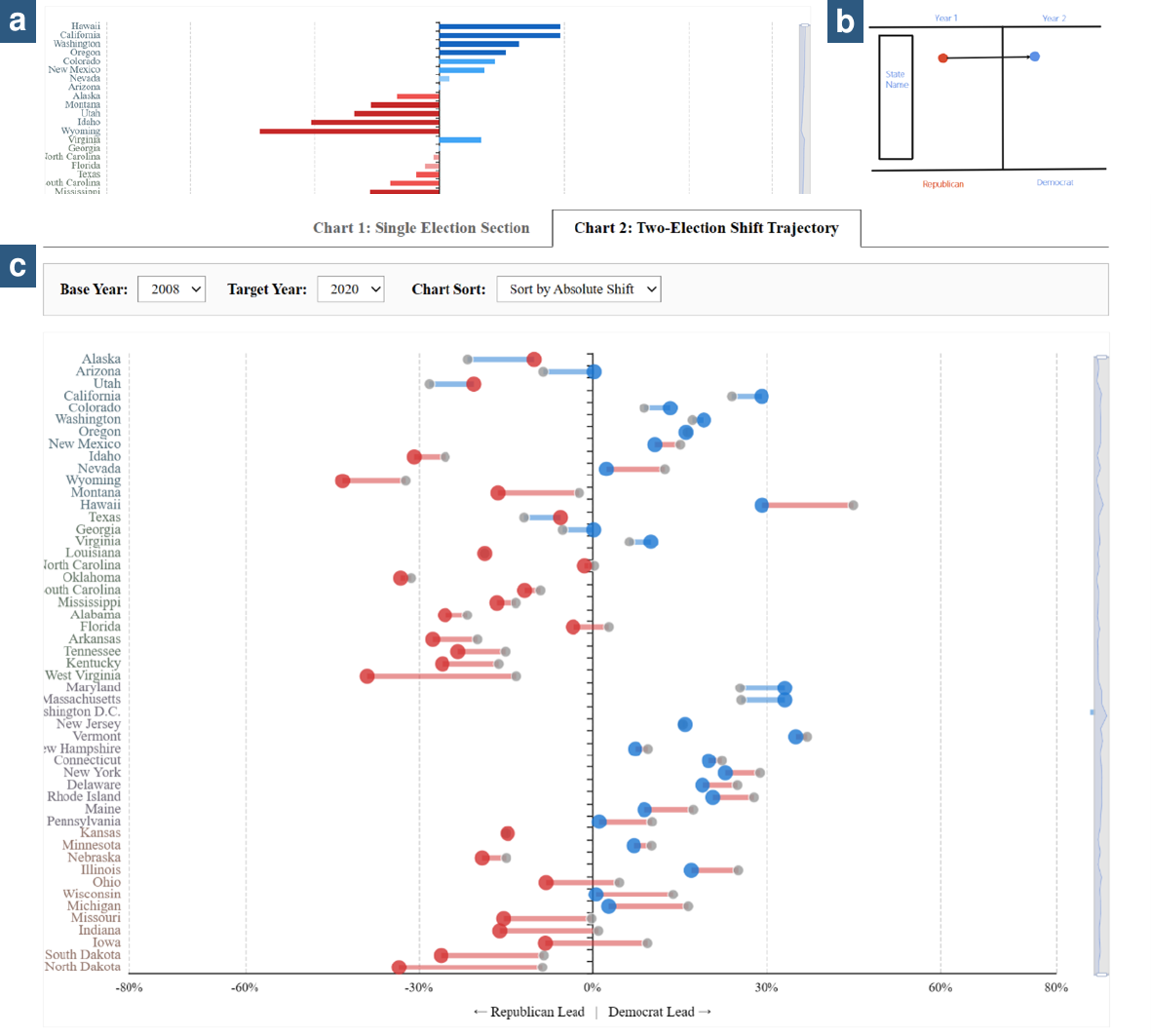}
 \caption{The co-design evolution of the \textit{U.S. Presidential Election Political Landscape Explorer} by P21. (a) Agent-generated bar chart; (b) User's conceptual sketch of a two-point design; (c) Final interactive dumbbell plot with data comparison between two selected years.}
 \label{fig:case_study_P21}
\end{figure}

\textbf{Human-agent collaboration for visualization design.} 
P21 sought to analyze and visualize the state-wise distribution of U.S. bipartisan votes over time. 
During the initial design phase, both P21 and the agent were constrained by the path dependence of traditional choropleth maps, which caused a creative impasse. 
The breakthrough came from a bidirectional bar chart generated by the agent (\cref{fig:case_study_P21}(a)). 
Serving as a vital visual stimulus, the chart prompted P21 to shift focus from geographic representations to numerical distributions, subsequently leading to a proposal to use bilateral scatter plots. 
The design logic was further refined when the agent suggested that ``colors could be instantaneously re-rendered across years to highlight partisan flips''. 
This inspired P21 to compare only two years in one visualization. 
When P21 attempted to express the changes through two-point connected arrows (\cref{fig:case_study_P21}(b)), but encountered a logical bottleneck, the agent leveraged its domain knowledge to introduce the dumbbell chart paradigm, ingeniously interpreting the length of the connecting lines as the degree of political swing, and finally completed the construction (\cref{fig:case_study_P21}(c)). 
This human-AI co-creation process demonstrates that in visualization design, the agent is not merely an executor of instructions but also a catalyst for human inspiration. 
By bridging technical and domain knowledge gaps, the agent facilitates a transition from rudimentary concepts to sophisticated, professional-grade visual narratives.

\section{Discussion}

\subsection{Design Implications}

Based on our analysis, we have identified shortcomings in current vibe coding tools regarding layout adjustments, interaction debugging, and visual version control. We distill the following design implications for vibe coding tools built for interactive visualizations:

\textbf{Multimodal detail edits through visual anchoring.} Allow users to select, annotate, drag, or adjust specific visual elements directly within the preview interface, and link the actions to underlying code, parameters, and natural language commands. Work such as DynaVis~\cite{dynavis:2024:chi} and DirectGPT~\cite{directgpt:2024:chi}, combining natural language with graphical interface editing, highlights the value of this approach for next-generation visual development tools.

\textbf{Visual debugging of interactive flows.} Visually trace data mappings, state propagation, and event sequences. By surfacing interaction histories and modification scopes, developers can efficiently debug issues in synchronization, event binding, and rendering logic.

\textbf{Controlled visual iterative collaboration.} Features such as local locking, difference previews, version rollback and design intent tracking enable developers to clearly specify which components, styles, interaction rules, or data processing logic can be modified by agents, and which elements must remain unchanged.

\subsection{Limitations and Future Work}

Our findings should be considered in light of several limitations. 
First, our participants were students majoring in computer science; therefore, the findings could not be generalized to users from other background. 
Second, since we did not standardize the models, platforms, or tools used by the participants, the performance of different models may have influenced the evaluation results. 
Finally, our study mainly focuses on the initial development process of interactive visualizations and does not yet cover subsequent secondary development and optimization. 
Future research can expand the sample pool to include participants with diverse professional backgrounds, development experience, and usage contexts. 
Long-term follow-up studies and periodic interviews could evaluate the performance of agent-generated code in terms of maintainability, scalability, and stability, and further analyze the division of labor and patterns of human-agent collaboration over time.

\section{Conclusion}

We employ an empirical study combining quantitative and qualitative methods to evaluate how developers use vibe coding to construct interactive visualizations. 
The study indicates that vibe coding significantly improves the prototyping efficiency of interactive visualization construction, particularly for generating static charts and simple interactions; however, it still has notable limitations regarding complex interactions, multi-view synchronization, visual details, and code stability. 
Vibe coding reshapes the development paradigm by shifting part of the coding burden towards semantic alignment and validation loops, resulting in noticeable friction during modal switching. 
Developers generally adopt a layered collaboration strategy, compensating for the lack of domain knowledge in the agent through task decomposition and selective manual code refinement, based on considerations of efficiency, complexity, or cognitive load. 
In summary, while vibe coding reduces the barrier to entry, the construction of high-quality systems still relies on precise human intervention. 
Future studies should focus on bridging the gap between natural language instructions and intuitive visual feedback to streamline the iterative design process.

\bibliographystyle{abbrv-doi}

\bibliography{template}

@article{sarkar2025vibe,
  title={Vibe coding: programming through conversation with artificial intelligence},
  author={Sarkar, Advait and Drosos, Ian},
  journal={arXiv preprint arXiv:2506.23253},
  year={2025}
}

@article{satyanarayan2016vega,
  title={Vega-lite: A grammar of interactive graphics},
  author={Satyanarayan, Arvind and Moritz, Dominik and Wongsuphasawat, Kanit and Heer, Jeffrey},
  journal={IEEE transactions on visualization and computer graphics},
  volume={23},
  number={1},
  pages={341--350},
  year={2016},
  publisher={IEEE}
}

@inproceedings{al-khalifa-2026-code-centric,
    title = "From Code-Centric to Concept-Centric: Teaching {NLP} with {LLM}-Assisted ``Vibe Coding''",
    author = "Al-Khalifa, Hend",
    booktitle = "In Proceedings of the Workshop on Teaching Natural Language Processing ({T}each{NLP} 2026)",
    month = mar,
    year = "2026",
    url = "https://aclanthology.org/2026.teachingnlp-1.3/",
    doi = "10.18653/v1/2026.teachingnlp-1.3",
    pages = "11--18",
    ISBN = "979-8-89176-375-3"
}

@article{D3tvcg2011,
author = {Bostock, Michael and Ogievetsky, Vadim and Heer, Jeffrey},
title = {D3 Data-Driven Documents},
year = {2011},
publisher = {IEEE Educational Activities Department},
address = {USA},
volume = {17},
number = {12},
issn = {1077-2626},
url = {https://doi.org/10.1109/TVCG.2011.185},
doi = {10.1109/TVCG.2011.185},
journal = {IEEE Transactions on Visualization and Computer Graphics},
pages = {2301–2309},
numpages = {9}
}

@INPROCEEDINGS{smartmlvs:2025:tvcg,
  author={Qiu, Tian and Wang, Fen and Huang, Shaohua and Guo, Meng and Zhao, Yuheng and Li, Jincheng and Chen, Siming},
  booktitle={Proceedings of IEEE Pacific Visualization Conference (PacificVis)}, 
  title={{SmartMLVs}: LLM-enabled Multiple Linked Views Generation for Interactive Visualization}, 
  year={2025},
  volume={},
  number={},
  pages={58-68},
  doi={10.1109/PacificVis64226.2025.00012}}

@ARTICLE{deepvis:2026:tvcg,
  author={Shuai, Zhihao and Li, Boyan and Yan, Siyu and Luo, Yuyu and Yang, Weikai},
  journal={IEEE Transactions on Visualization and Computer Graphics}, 
  title={DeepVIS: Bridging Natural Language and Data Visualization Through Step-Wise Reasoning}, 
  year={2026},
  volume={32},
  number={1},
  pages={868-878},
  doi={10.1109/TVCG.2025.3634645}}

@inproceedings{dynavis:2024:chi,
author = {Vaithilingam, Priyan and Glassman, Elena L. and Inala, Jeevana Priya and Wang, Chenglong},
title = {{DynaVis}: Dynamically Synthesized UI Widgets for Visualization Editing},
year = {2024},
isbn = {9798400703300},
url = {https://doi.org/10.1145/3613904.3642639},
doi = {10.1145/3613904.3642639},
booktitle = {Proceedings of the CHI Conference on Human Factors in Computing Systems},
articleno = {985},
numpages = {17},
location = {Honolulu, HI, USA}
}

@inproceedings{directgpt:2024:chi,
author = {Masson, Damien and Malacria, Sylvain and Casiez, G\'{e}ry and Vogel, Daniel},
title = {{DirectGPT}: A Direct Manipulation Interface to Interact with Large Language Models},
year = {2024},
isbn = {9798400703300},
publisher = {Association for Computing Machinery},
address = {New York, NY, USA},
url = {https://doi.org/10.1145/3613904.3642462},
doi = {10.1145/3613904.3642462},
booktitle = {Proceedings of the CHI Conference on Human Factors in Computing Systems},
articleno = {975},
numpages = {16},
location = {Honolulu, HI, USA},
series = {CHI '24}
}

@ARTICLE{chartgpt:2025:tvcg,
  author={Tian, Yuan and Cui, Weiwei and Deng, Dazhen and Yi, Xinjing and Yang, Yurun and Zhang, Haidong and Wu, Yingcai},
  journal={IEEE Transactions on Visualization and Computer Graphics}, 
  title={ChartGPT: Leveraging LLMs to Generate Charts From Abstract Natural Language}, 
  year={2025},
  volume={31},
  number={3},
  pages={1731-1745},
  doi={10.1109/TVCG.2024.3368621}}

@ARTICLE{vibe-coding-survey:2025:arxiv,
      title={A Survey of Vibe Coding with Large Language Models}, 
      author={Yuyao Ge and Lingrui Mei and Zenghao Duan and Tianhao Li and Yujia Zheng and Yiwei Wang and Lexin Wang and Jiayu Yao and Tianyu Liu and Yujun Cai and Baolong Bi and Fangda Guo and Jiafeng Guo and Shenghua Liu and Xueqi Cheng},
      year={2025},
      eprint={2510.12399},
      archivePrefix={arXiv},
      primaryClass={cs.AI},
      journal = {preprint arXiv:2510.12399}
}

@ARTICLE{vega-lite:2017:tvcg,
  author={Satyanarayan, Arvind and Moritz, Dominik and Wongsuphasawat, Kanit and Heer, Jeffrey},
  journal={IEEE Transactions on Visualization and Computer Graphics}, 
  title={Vega-Lite: A Grammar of Interactive Graphics}, 
  year={2017},
  volume={23},
  number={1},
  pages={341-350},
  doi={10.1109/TVCG.2016.2599030}}

@inproceedings{declarative:2014:uist,
author = {Satyanarayan, Arvind and Wongsuphasawat, Kanit and Heer, Jeffrey},
title = {Declarative interaction design for data visualization},
year = {2014},
isbn = {9781450330695},
url = {https://doi.org/10.1145/2642918.2647360},
doi = {10.1145/2642918.2647360},
booktitle = {Proceedings of the Annual ACM Symposium on User Interface Software and Technology},
pages = {669–678},
numpages = {10},
location = {Honolulu, Hawaii, USA},
series = {UIST '14}
}

@ARTICLE{nested-model:2009:tvcg,
  author={Munzner, Tamara},
  journal={IEEE Transactions on Visualization and Computer Graphics}, 
  title={A Nested Model for Visualization Design and Validation}, 
  year={2009},
  volume={15},
  number={6},
  pages={921-928},
  doi={10.1109/TVCG.2009.111}}

@inproceedings{copilot-user-behavior:2024:chi,
author = {Mozannar, Hussein and Bansal, Gagan and Fourney, Adam and Horvitz, Eric},
title = {Reading Between the Lines: Modeling User Behavior and Costs in AI-Assisted Programming},
year = {2024},
isbn = {9798400703300},
url = {https://doi.org/10.1145/3613904.3641936},
doi = {10.1145/3613904.3641936},
booktitle = {Proceedings of the CHI Conference on Human Factors in Computing Systems},
articleno = {142},
numpages = {16},
location = {Honolulu, HI, USA},
series = {CHI '24}
}

@inproceedings{validate-ai-code:2024:chi,
author = {Ferdowsi, Kasra and Huang, Ruanqianqian (Lisa) and James, Michael B. and Polikarpova, Nadia and Lerner, Sorin},
title = {Validating AI-Generated Code with Live Programming},
year = {2024},
isbn = {9798400703300},
url = {https://doi.org/10.1145/3613904.3642495},
doi = {10.1145/3613904.3642495},
booktitle = {Proceedings of the CHI Conference on Human Factors in Computing Systems},
articleno = {143},
numpages = {8},
location = {Honolulu, HI, USA},
series = {CHI '24}
}

@ARTICLE{lightva:2025:tvcg,
  author={Zhao, Yuheng and Wang, Junjie and Xiang, Linbing and Zhang, Xiaowen and Guo, Zifei and Turkay, Cagatay and Zhang, Yu and Chen, Siming},
  journal={IEEE Transactions on Visualization and Computer Graphics}, 
  title={{LightVA}: Lightweight Visual Analytics With LLM Agent-Based Task Planning and Execution}, 
  year={2025},
  volume={31},
  number={9},
  pages={6162-6177},
  doi={10.1109/TVCG.2024.3496112}}

@inproceedings{celestial:2026:chi,
  author       = {Priyan Vaithilingam and
                  Alan Leung and
                  Jeffrey Nichols and
                  Titus Barik},
  title        = {The Way We Notice, That's What Really Matters: Instantiating {UI} Components with Distinguishing Variations},
  booktitle    = {Proceedings of the {CHI} Conference on Human Factors in Computing Systems},
  pages        = {900:1--900:18},
  publisher    = {{ACM}},
  year         = {2026},
  url          = {https://doi.org/10.1145/3772318.3790621},
  doi          = {10.1145/3772318.3790621},
  bibsource    = {dblp computer science bibliography, https://dblp.org}
}

@ARTICLE{llm-vis-item:2025:tvcg,
  author={Cui, Yuan and Ge, Lily W. and Ding, Yiren and Harrison, Lane and Yang, Fumeng and Kay, Matthew},
  journal={IEEE Transactions on Visualization and Computer Graphics}, 
  title={Promises and Pitfalls: Using Large Language Models to Generate Visualization Items}, 
  year={2025},
  volume={31},
  number={1},
  pages={1094-1104},
  doi={10.1109/TVCG.2024.3456309}}

@ARTICLE{data-formulator:2024:tvcg,
  author={Wang, Chenglong and Thompson, John and Lee, Bongshin},
  journal={IEEE Transactions on Visualization and Computer Graphics}, 
  title={Data Formulator: AI-Powered Concept-Driven Visualization Authoring}, 
  year={2024},
  volume={30},
  number={1},
  pages={1128-1138},
  doi={10.1109/TVCG.2023.3326585}}

@misc{propose-vibe-coding:2025:x,
  author = {Karpathy, Andrej},
  title = {There's a new kind of coding {I} call ``vibe coding''...},
  year = {2025},
  howpublished = {Tweet on X (formerly Twitter)},
  url = {https://x.com/karpathy/status/1886192184808149383},
  note = {Available: https://x.com/karpathy/status/1886192184808149383 [Accessed: Apr. 1, 2026]}
}

@misc{gemini:2026:google,
  author       = {{Google}},
  title        = {{Gemini 3.1 Pro Preview}},
  year         = {2026},
  howpublished = {Google AI Studio},
  url          = {https://aistudio.google.com},
  note         = {Available: https://aistudio.google.com [Accessed: Apr. 12, 2026]}
}
\end{document}